\documentclass{ws-procs961x669}

\begin{document}


\title{Scalar Perturbations of Regular Black Holes\\ derived from a Non-Singular Collapse Model in Asymptotic Safety}

\author{A. Spina$^{a,b}$, S. Silveravalle$^{d,e,f}$ and A. Bonanno$^{b,c}$}

\address{$^a$Department of physics and astronomy, University of Catania, Via S. Sofia 64, 95123 Catania, Italy\\
$^b$INFN, Sezione di Catania, Via S. Sofia 64, 95123 Catania, Italy\\
$^c$INAF, Osservatorio Astrofisico di Catania, Via S. Sofia 78, 95123 Catania, Italy\\
$^{d}$SISSA - International School for Advanced Studies, Via Bonomea 265, 34136 Trieste, Italy\\
$^{e}$INFN, Sezione di Trieste, Via Valerio 2, 34127 Trieste, Italy\\
$^{f}$IFPU - Institute for Fundamental Physics of the Universe, Via Beirut 2, 34151 Trieste, Italy
Contact E-mail: Andrea.Spina@phd.unict.it\\}

\begin{abstract}
We investigate the massless scalar field perturbations, focusing on the quasinormal modes spectrum and the ringdown waveform of regular black hole spacetimes derived within the Asymptotic Safety program. In particular, we discuss the stability of a new class of AS black holes recently derived dynamically within a non-singular model of collapse and explore the possibility of detecting signatures of the horizon structure with high-order overtones.
\end{abstract}

\keywords{Regular Black Holes, Quasinormal Modes, Stability}

\bodymatter

\section{Introduction}

In General Relativity black holes are among the most fascinating solutions to Einstein's equations, describing regions of spacetime with intense gravitational fields and unique properties. When isolated they are some of the simplest objects in physics, characterized only by few parameters, which are their mass, angular momentum and charge. However, these solutions, such as the Schwarzschild and Kerr metrics, are plagued by a fundamental issue: the presence of singularities, regions where the curvature of spacetime becomes infinite and classical physical laws of General Relativity break down. One of the most widely explored approaches to address the singularity problem is the concept of regular black holes \cite{Lan_2023}. These black hole solutions have non-singular cores thanks to a direct modification of the spacetime geometry. While several different proposals exist regarding the exact form and formation mechanism of regular black holes \cite{Bonanno:2000ep, Hayward:2005gi, Dymnikova:1992ux}, they all share the common goal of eliminating singularities through these modifications.
Although regular black holes solve the problem of singularities, they present new challenges. One of the most pressing issues is how to distinguish these non-singular black holes from their classical singular counterparts, as well as from one other.

In this context, studying perturbations of black holes becomes a powerful tool. When a black hole is perturbed, whether by a passing object or during a merger, the spacetime responds by emitting waves that decay over time. These decaying signals, known as ringdown waveforms, carry information about the structure of the black hole, the nature of the perturbations and the surrounding spacetime. 
The quasinormal modes (QNM), which are the characteristic complex frequencies of these oscillating black holes, play a central role in the ringdown phase \cite{Berti_2009}. By studying these QNM we can learn about the stability of the black hole, as well as differentiate between black holes arising from different solutions (as in the case of various regular black holes). The study of perturbations and quasinormal modes appears to be the most effective method for testing solutions and deviations from classical models, particularly through the analysis of gravitational waves \cite{Franchini:2023eda}. This approach is especially advantageous when compared to investigations into black hole shadows conducted by the Event Horizon Telescope (EHT), which has a $10\%$ margin of error in the observed dimension of the shadow \cite{2019ApJ...875L...1E} and currently provides information only about supermassive black holes. In contrast, just by analyzing only the fundamental mode of gravitational waves we already achieve over $90\%$ confidence in the parameter values \cite{Carullo_2019, Isi_2019}. Furthermore, with future detectors and the inclusion of overtone modes, we will have a significant increase in precision \cite{Carullo_2019, Isi_2019}.

In this work we focus on a specific regular black hole model that was recently proposed \cite{PhysRevLett.132.031401}. It is based on the description of dynamically collapsing matter, and is particularly intriguing because it directly predicts the formation of a regular black hole during the collapse process rather than assuming these objects to be static or relying on exotic matter fields. The model operates under the premise that black hole solutions observed in nature are generated by an interior matter configuration whose evolution remains free of singularities, owing to the antiscreening behavior of the running gravitational constant at small scales. This behavior can be obtained from the Asymptotic Safety (AS) framework \cite{10.21468/SciPostPhys.12.1.001}, which aims to provide a consistent theory of quantum gravity across all energy scales. AS is built on the idea that the gravitational constant becomes "safe" at high energies, avoiding unphysical divergences. This is achieved through a non-trivial UV fixed point in the renormalization group flow, at which the couplings of the theory converge to finite values \cite{Reuter_1998} and physical observables remain finite. Furthermore, this model is of particular interest because the antiscreening mechanism is implemented starting from an effective Lagrangian coupled to the matter, distinguishing it from many other models.

In the present study we investigate the scalar field perturbations of this non-singular black hole model, focusing on its stability through the analysis of quasinormal modes and the ringdown waveform. Specifically, we will calculate the QNM frequencies which have been explored recently \cite{Stashko:2024wuq}; however, our investigation extends to higher overtones and to the comparison with the classical Schwarzschild black hole and other regular black hole solutions. Additionally, we analyze the full time evolution of the ringdown waveform through numerical integration, and make comparison with various regular black holes solutions. Through this comparative analysis, we aim to identify key differences in the behavior of the perturbations, particularly at higher overtones where distinctions between the different solutions become more pronounced.

\section{Non-singular collapse model}

Let us proceed into more detail in the description of the previously introduced model. To illustrate this regular black hole, we can divide the analysis into two regions: the interior where we have the collapsing matter cloud, and a static exterior part where the event horizon is present. We note that the study of scalar perturbation only needs the exterior metric, because all observables come from the event horizon onwards. 
On the other hand, the interior metric carries all the information on the regular core, and the match between the two regions will determine the horizon structure and will be reflected indirectly in the external measurements of the perturbations.

Following the approach of Markov and Mukhanov \cite{Markov1985DeSI} we start with the action
\begin{equation}
    S=\frac{1}{16{\pi}G_N}\int{d^{4}x\sqrt{-g}[R+2\chi(\epsilon)\mathcal{L}]},
\end{equation}
where $\chi$ represents a gravity-matter coupling, which is a function of the proper energy density $\epsilon$ of the matter fluid. From the total variation of the action, we can obtain the following field equations \cite{PhysRevLett.132.031401}:
\begin{equation}
    R_{\mu\nu}-\frac{1}{2}g_{\mu\nu}R=\frac{\partial(\chi\epsilon)}{\partial\epsilon}T_{\mu\nu}+\frac{\partial(\chi)}{\partial\epsilon}\epsilon^{2}g_{\mu\nu},
\end{equation}
that define a relation between the coupling $\chi$ and an effective gravitational constant \cite{PhysRevLett.132.031401} which also depends on the proper energy density
\begin{equation}
    8{\pi}G(\epsilon)\equiv\frac{\partial(\chi\epsilon)}{\partial\epsilon}.
\end{equation}
Now, considering for the interior a FLRW metric that depends on the scale factor $a(t)$,
if we assign $G(\epsilon)$, from the previous equation we can obtain $\chi(\epsilon)$ and have all the information from the lagrangian and of the interior. 
To determine the behaviour of $G(\epsilon)$, we can refer to the functional renormalization group in the context of the asymptotic safety (AS) program.

Following Ref.~\citenum{10.21468/SciPostPhys.12.1.001}, we adopt the running gravitational constant $G(k)$ where $k$ is the momentum scale, and under the assumption of dust collapse, we can connect $k$ with $\epsilon$ expressing our $G(\epsilon)$ as \cite{PhysRevLett.132.031401}
\begin{equation}
    G(\epsilon)=\frac{G_N}{1+\xi\epsilon}
\end{equation}
where $\xi$ is a characteristic scale of the system that was introduced. This represents the primary parameter distinguishing our model from classical cases.

\begin{figure}[htb]
\begin{center}
    \includegraphics[width=3.1in]{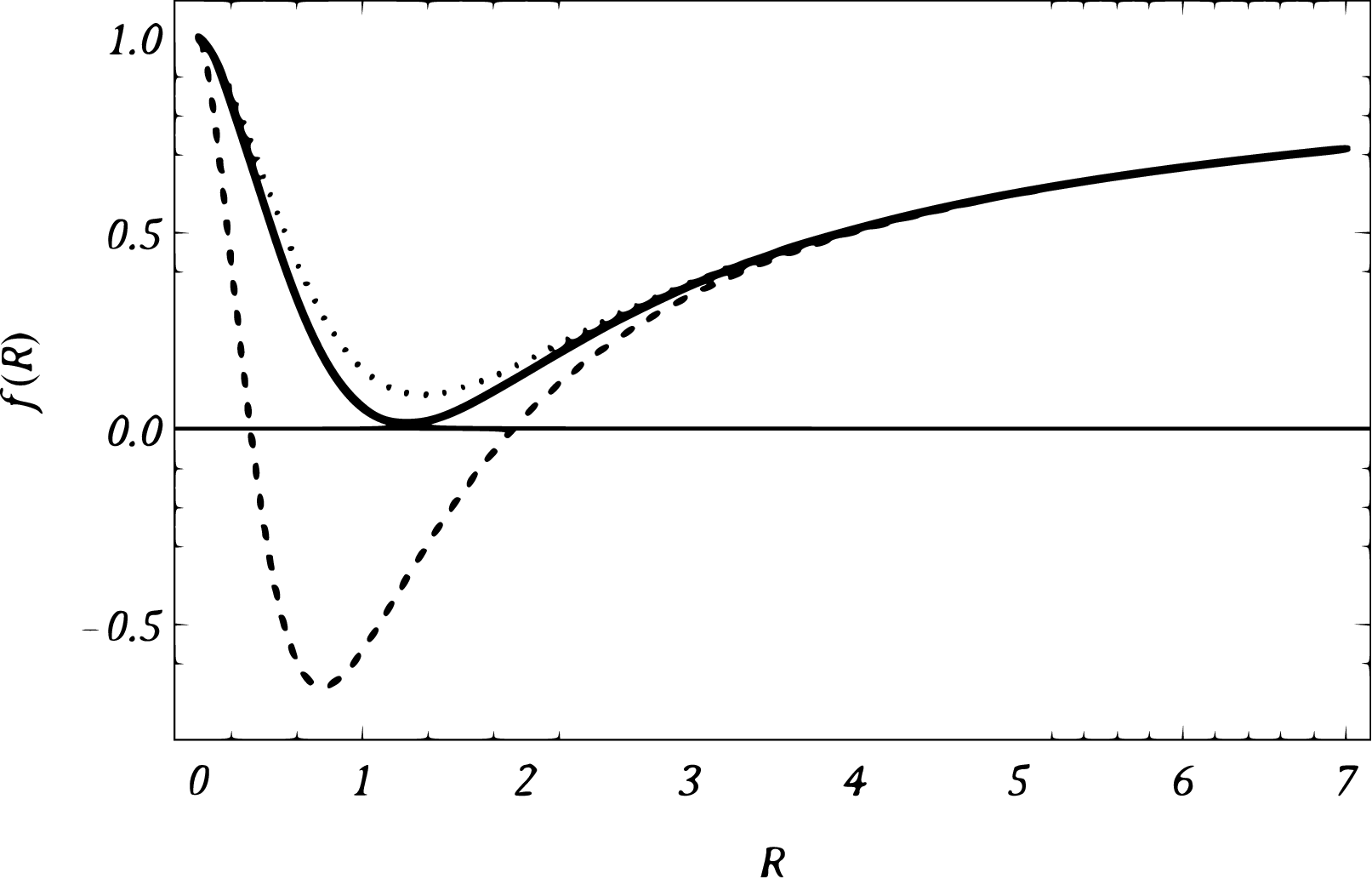}
    \end{center}
    \caption{Shape of $f(r)$ with different values of $\xi$ and $M_0=1$. For $\xi<\xi_{cr}$ dashed line, for $\xi=\xi_{cr}$ thick line, for $\xi>\xi_{cr}$ dotted line }
    \label{fig0}
\end{figure}

With this choice of $G(\epsilon)$ we can obtain also the behavior of the scale factor at large times \cite{PhysRevLett.132.031401}
\begin{equation}
    a(t)\simeq{e^{-t^{2}/4\xi}}
\end{equation}
that we note depends on the parameter $\xi$.
At this stage, we are ready to implement the matching between the collapsing matter cloud and the exterior system.\\
By considering a generic static and spherically symmetric metric (imposing from now on $c=G_N=\hbar=1)$
\begin{equation}
  ds^2=-f(r)dT^2+\frac{1}{f(r)}dr^2+{r^2}d\Omega^2  
\end{equation}
where 
\begin{equation}
    f(r)=1-\frac{2M(r)}{r}
\end{equation}
and imposing a continuous matching at a finite boundary between the two regions, that depends on the scale factor $a(t)$ \cite{PhysRevLett.132.031401}, we can derive an expression for $M(r)$ \cite{PhysRevLett.132.031401} and consequently, the corresponding shape of $f(r)$ as:
\begin{equation}
    f(r)=1-\frac{r^2}{3\xi}\log{(1+\frac{6M_0\xi}{r^3})}. \label{f(r)}
\end{equation}
There is $\xi_{cr}$, that is the critical value of the characteristic scale for which the transcendental equation $f(r)=0$ has two coinciding solution, that means there are two coinciding horizons and the black hole became extremal. This discriminates the number of solutions for the existence of zero, one, or two horizons as illustrated in Fig \ref{fig0}.

Another important point is that $f(r)$ is defined only for positive $r$, ensuring that it never diverges.

Equation (\ref{f(r)}) represents the main result that will be used to study the perturbations in the following sections.
\section{Scalar perturbations of black holes}

Our focus is on investigating the scalar field perturbations for the non-singular black hole model discussed earlier. Scalar perturbations are particularly useful in this context as they provide a simplified yet effective framework for analyzing the dynamical response of black holes to external disturbances.

By studying the quasinormal mode (QNM) frequencies and the resulting ringdown waveform, we can evaluate the stability of the black hole and extract key features that allow us offers a method for distinguishing between different black hole based on their QNM spectra.

To analyze the perturbation we start from Einstein equation in the presence of a scalar field, in the linear approximation, the equations reduce to the propagation of the scalar field on the background metric of the black hole. Note that the equation of scalar perturbations could represents also axial components of the gravitational perturbations \cite{Nollert:1999ji}, as consequence they gives information also on them. For our case, this corresponds to the equation of motion of a massless scalar field, which reduces to the massless Klein-Gordon equation in a curved spacetime \cite{Konoplya_2011}.
\begin{equation}
    \frac{1}{\sqrt{-g}}\partial_\mu(\sqrt{-g}g^{\mu\nu}\partial_\nu\psi)=0 \label{eqKG}
\end{equation}
The solution of this equation provides the massless scalar perturbation for our generic metric. 
In the case of spherical symmetry we can express the ansatz for the wave function as 
\begin{equation}
    \psi=Y_{lm}(\theta,\varphi)\phi(t,r)/r  \label{ansaltz}.
\end{equation}
Substituting these conditions into equation (\ref{eqKG}), and introducing the tortoise coordinate $dr_*=dr/f(r)$, yelds the following form:

\begin{equation}
    \left(\frac{d^2}{dt^2}-\frac{d^2}{dr^2_*}\right)\phi(r,t)+V(r)\phi(r,t)=0  \label{eq_t}.
\end{equation}

\begin{figure}[htb]
\begin{center}
    \includegraphics[width=3.5in]{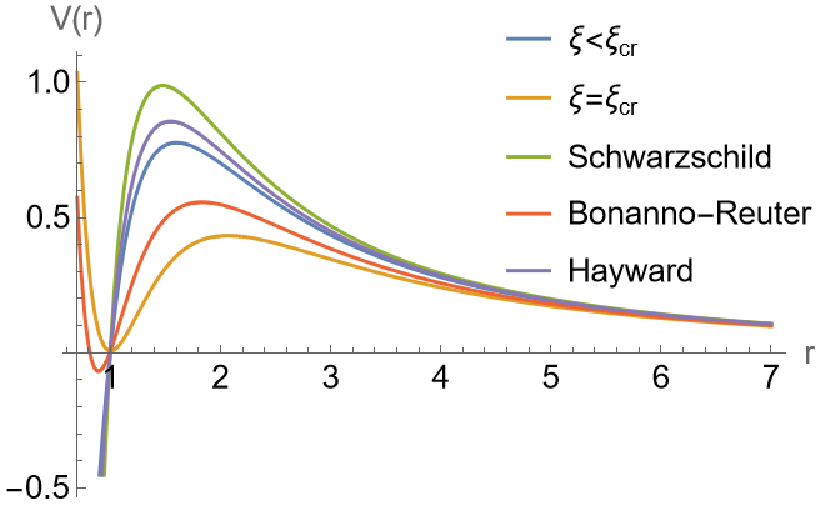}
    \end{center}
    \caption{Plot of the potential with same horizon radius and with $l=2$ of: the non singular collapse model (with different values of $\xi$), Schwarzschild and other regular black holes we will compare in next sections.}
    \label{pot}
\end{figure}

If we now assume a decomposition into frequency modes, we express the time dependence as
\begin{equation}
    \phi(t,r)=e^{-i{\omega}t}\Gamma(r)
\end{equation}
where $\omega$ can have both real and imaginary parts. The real part of $\omega$ represents the oscillation frequency, while the imaginary part determines the damping rate. These $\omega$ values, both real and imaginary, correspond to the quasinormal modes of the system, which will be the key quantities we analyze.
Substituting the previous decomposition in the eq.\ref{eq_t}, we will obtain the following equation:

\begin{equation}
     -\frac{d^2\Gamma}{dr^2_*}+V(r)\Gamma(r)=\omega^2\Gamma(r) \label{master_eq}
\end{equation}

with a potential $V(r)$ which has the form :
\begin{equation}
    V(r)=f(r)\left(\frac{l(l+1)}{r^2}+\frac{f'(r)}{r}\right)
\end{equation}
where l is  the multipole quantum number, that gives the angular distribution. (We show some of the potential that next we will use in fig.\ref{pot})

At this stage, eq. \eqref{master_eq}, with a specific potential, will be the master equation that we need to solve.

\subsection{Quasinormal modes and stability}

From the Eq.(\ref{master_eq}), we seek to obtain the well-known quasinormal modes (QNM). These are defined as solutions to Eq.(\ref{master_eq}) with specific boundary conditions imposed at the event horizon and at spatial infinity \cite{Berti_2009}. At the event horizon, according to classical theory, nothing should escape the event horizon: only ingoing modes should exist, and thus the boundary condition is:
$$\phi\to{e^{-i{\omega}(t+r_*)}},~~r_*\xrightarrow{} -\infty,$$
while at spatial infinity, nothing should come from infinity entering in the spacetime, so the condition is: 
$$\phi\to{e^{-i{\omega}(t-r_*)}},~~r_*\xrightarrow{} +\infty.$$

By studying the QNMs, we can investigate the stability of the objects described by our metric. As said before quasinormal modes are complex frequencies where the real part represents the oscillation frequency, while the imaginary part dictates the rate of decay or growth of the oscillation.
Specifically, if the imaginary part is negative, the perturbation decays over time, indicating that the system is stable. On the other hand, a positive imaginary part would imply an instability, as the perturbation would grow with time. Thus, the sign of the imaginary part is crucial for determining stability.
\section{QNM computation and comparison (WKB method)}

To obtain the frequencies of QNM we employ the WKB method, originally developed by Will and Schutz \cite{1985ApJ...291L..33S} and later extended to higher order of approximation \cite{Konoplya:2004ip}. The core idea of this method is to divide the spacetime into three regions: two near the boundaries (infinities) and one central region where the potential has its peak (see Fig.1 of Ref.~\citenum{Konoplya_2011} In this central zone, an exponential expansion is applied that depends on the potential itself. By imposing the appropriate boundary condition and connecting the solution across these areas, the QNM frequencies can be determined. In particular, at higher orders, the formula for the frequencies is given by \cite{Konoplya:2019hlu}:
\begin{equation}
\begin{split}
    \omega^2 =&\, V_0 + A_2(K^2) + A_4(K^2) + A_6(K^2) +...\\
    &- iK\sqrt{- 2V_2}\left(1 + A_3(K^2) + A_5(K^2) + A_7(K^2)+...\right)
\end{split}
\end{equation}
where $K$ takes halfinteger values, $V_i$ represents the values of higher-order derivatives of the potential and $A_i$ are polynomials of $V_i$.

Now, we move to the calculation of the QNM for the model introduced in Sec.1, with $n=l$, for better stability of the WKB computation, where $n$ in the overtones number, that physically correspond to the excited modes, with a more negative imaginary part of the frequencies and a smaller real part, of the black hole. Therefore, higher overtones represent the faster decaying components of the signal and they reveal more subtle differences in the black hole’s response to perturbations, allowing us to distinguish solution with greater clarity. \cite{Konoplya_2022}.

\begin{figure}[htb]
\begin{center}
    \includegraphics[width=3.1in]{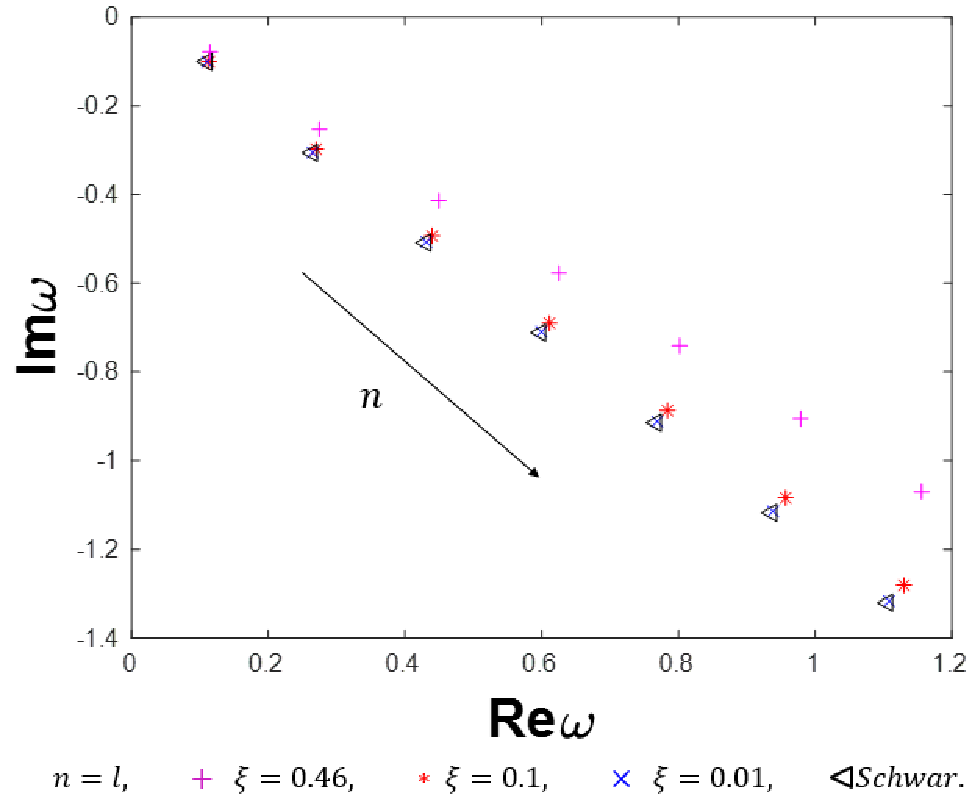}
    \end{center}
    \caption{Comparison of real and imaginary part of QNM frequencies, increasing the overtones $n$, for black holes with mass $M=1$  and different values of $\xi$ with Schwarzschild.}
    \label{fig1}
\end{figure}

First we made a comparison, between different values of the parameter $\xi$ ($\xi=0.1$, $\xi=0.01$, $\xi=\xi_{cr}=0.46$) for the non singular model with those of the Schwarzschild BH to observe how the parameter affects the deviation from the classical case (see Fig.\ref{fig1}).

We observe that all the modes are stable, as the imaginary part of the frequencies are negative and as expected, the differences between the non-singular model and the Schwarzschild black hole increase with the parameter $\xi$. In fact, when we consider small values of $\xi$ (like $\xi=0.01$ in the graph), differences with Schwarzschild black hole are negligible. This trend becomes even more pronounced when considering higher overtones, denoted by the letter $n$ in the graph. Note also that our data are in agreement (with the differences between the values with a relative error less than $10^{-3}$) with those calculated in \cite{Stashko:2024wuq}.

\begin{figure}[htb]
\begin{center}
    \includegraphics[width=3.1in]{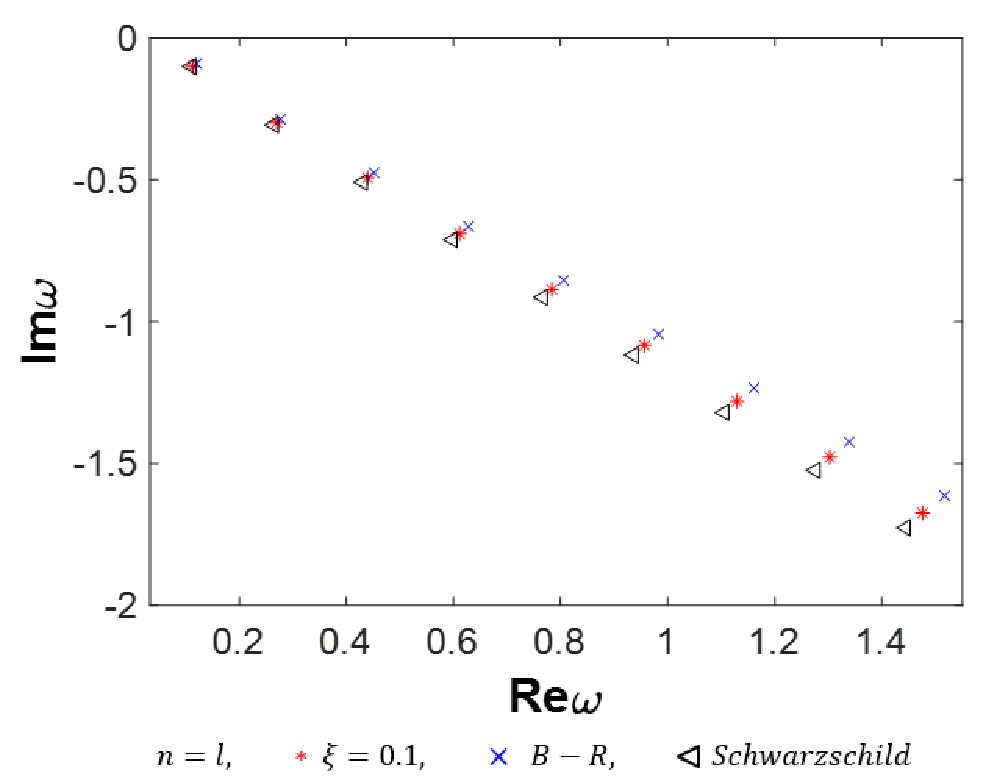}
    \end{center}
    \caption{QNM frequencies comparison increasing the overtones (going to the right),  with the same values of the mass of the black holes fixed to $M_0=1$.}
    \label{fig2a}
\end{figure}

\begin{figure}[htb]
\begin{center}
    \includegraphics[width=3.1in]{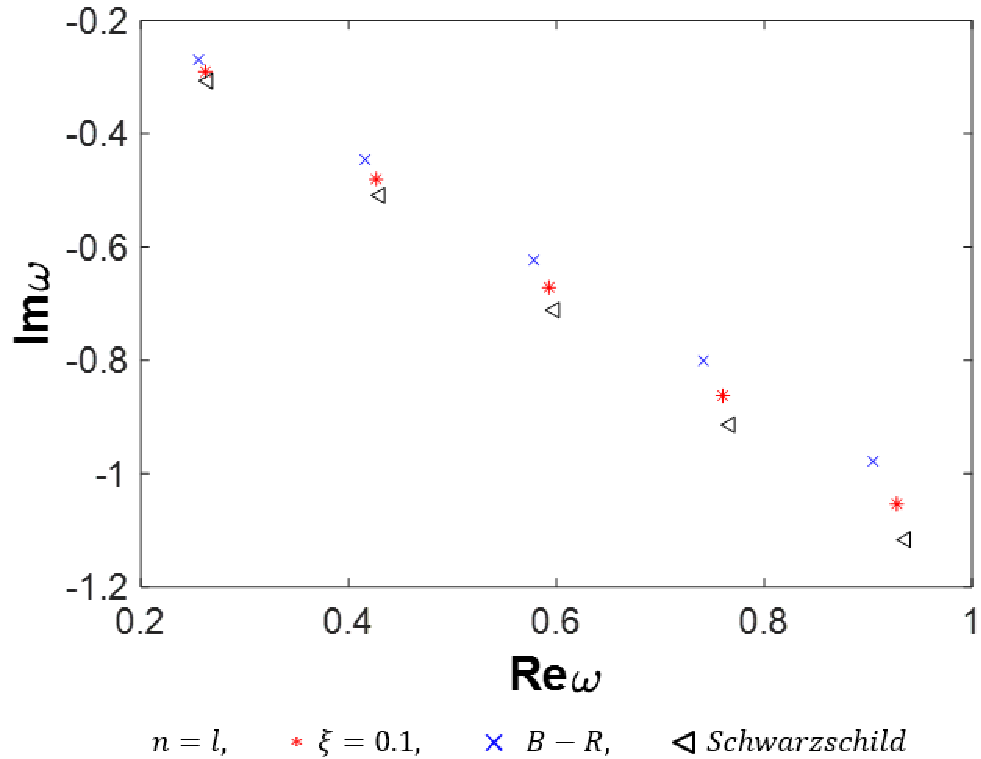}
    \end{center}
    \caption{QNM frequencies comparison increasing the overtones (going to the right), with the same horizon radius $r_h$ fixed at $r_h=1/2$.}
    \label{fig2b}
\end{figure}

Having demonstrated the stability of the model and how the frequencies vary with the parameter $\xi$, we extend the comparison beyond the Schwarzschild black hole by analyzing another regular black holes solution, specifically the one developed by Reuter and Bonanno \cite{Bonanno:2000ep}, which is an interesting static solution also derived from asymptotic safety, so an ideal comparison with the non-singular model. In particular, we examined the QNM frequencies for three models: the non-singular collapse black hole (with $\xi=0.1$), the Schwarzschild black hole, and the Reuter-Bonanno black hole. These comparisons are made under two conditions: first, with the same mass for all models (Fig. \ref{fig2a}), and second, with the same horizon radius (Fig. \ref{fig2b}), since the relationship between mass and horizon radius varies from one solution to another.

In both cases, we see that at higher overtones, the differences between the models become more significant, primarily due to the differences in horizon structure. Physically, the overtones increasing sensitivity to the geometry near the horizon, means that discrepancies between models become more evident at more negative imaginary part of frequencies, so for the more damped modes.
This suggests that while the fundamental modes are more similar, the higher overtones capture finer details of the black hole's structure. As a result, the modes between solutions are not uniformly suppressed.

We can also note that while at equal mass, the regular black holes exhibit both larger real and imaginary parts compared to Schwarzschild so they are more oscillating and less suppressed, at equal horizon radius, they show larger imaginary parts but smaller real parts. This indicates that, this time the modes are both less suppressed and oscillating.

Therefore, in principle, it is possible to differentiate between these models based on their QNM spectra, which is a key goal for testing and distinguishing different solutions.

\section{Ringdown Waveform}

To gain a broader understanding of stability and the differences between the models, we extend our analysis to the ringdown waveform in the time domain.

To this end, we begin from Eq. (\ref{eq_t}), to solve it numerically we transform it into light-cone coordinates $v=t+r_*$ and $u=t-r_*$, and discretizing the system on a grid to integrate the function as \cite{Silveravalle:2023lnl,Konoplya_2011}:
\begin{equation}
    \phi(N)=\phi(W)+\phi(E)-\phi(S)+\frac{h^2}{8}V(S)(\phi(W)+\phi(E))
\end{equation}
Where $N,W,E,S$ are adjacent point of the grid and $h$ is the step size.

\begin{figure}
\begin{center}
    \includegraphics[width=4.7in]{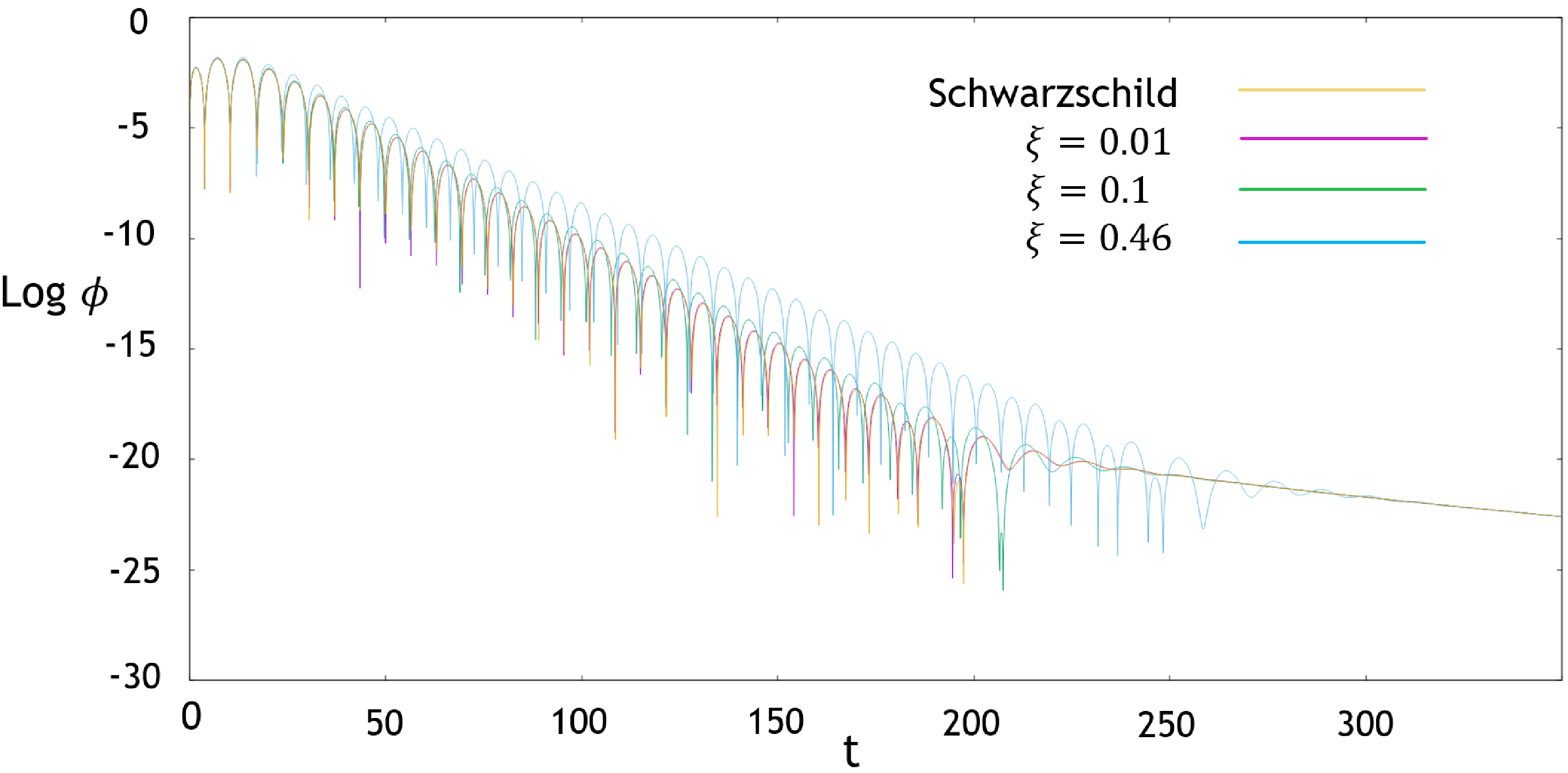}
    \end{center}
    \caption{Waveform comparison for the Non-Singular collapse model with $M=1$ and $l=2$ at different values of $\xi$ with Schwarzschild.}
    \label{fig3}
\end{figure}

As first step, we compare the ringdown waveforms for different values of the parameter $\xi$, including the critical value ($\xi=0.46$ in this case), to observe the behavior of the ringdown as it approaches this threshold  (Fig.\ref{fig3}) and compare them with the Schwarzschild case.

\begin{figure}
\begin{center}
    \includegraphics[width=4.7in]{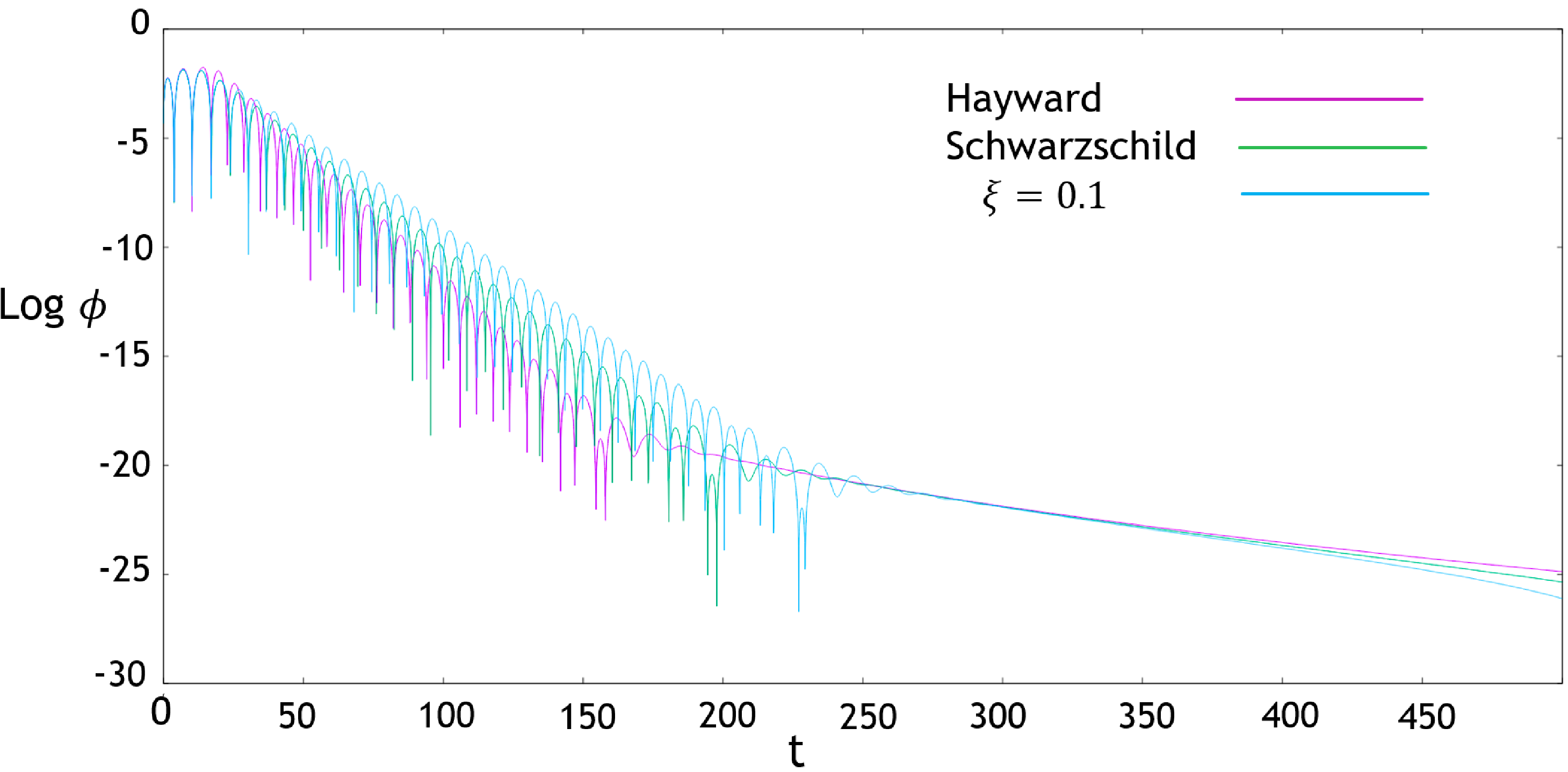}
    \end{center}
    \caption{Regular Black Holes Waveform comparison with Schwarzschild ($M=1$ and $l=2$).}
    \label{fig4}
\end{figure}

We then extend our comparison to the well-known Hayward regular black hole solution \cite{Hayward:2005gi}, which is one of the first developed and widely studied due to its simplicity in the context of regular black holes. As with the frequency analysis conducted earlier, we examine the behavior of the ringdown waveform for these solution and Schwarzschild (Fig. \ref{fig4}).

From the analysis of Fig. \ref{fig4}, we observe that, as expected, all modes remain stable. In the early stages of the ringdown, the waveforms of the three black holes are almost indistinguishable, as the fundamental mode dominates. However, as time progresses and the higher overtones become more significant, the differences between the solutions increase.

While this time-domain analysis may lose some of the quantitative precision provided by the frequency-based approach, it confirms the overall stability of the solutions, independent of the time dependence assumed in earlier calculations. Furthermore, it explicitly illustrates how the differences between black hole solutions grow with the increasing significance of higher overtones, enhancing our qualitative understanding of scalar field perturbations.

Specifically, it is explicitly observed that the ringdown decays at varying speeds overall, and this is of significant importance.
This behavior implies that, in principle, one could better distinguish between the black hole solutions through careful observation of the ringdown signal. 

\section{Conclusions}

In this work we have investigated the scalar perturbations of regular black hole solutions which arise from a non-singular collapse model in the context of asymptotically safe gravity, focusing on the quasinormal modes frequency spectrum and the time profile of the ringdown waveform. In particular we analyzed the impact of the key parameter $\xi$ which characterizes the running of the gravitational constant at high energies, and at the practical level distinguishes the regular black hole from the classical Schwarzschild solution, and compared this model with other regular black holes present in the literature.

We observed both in time and frequency domain that the model is stable against scalar perturbations for values of $\xi\in[0,\xi_c]$, where $\xi_c$ is the critical value for which the black holes become extremal. Regarding the QNM spectra we show that increasing values of $\xi$ give rise to more pronounced deviations from the Schwarzschild case, as expected. Nonetheless these deviations become more relevant at higher overtones: while for the fundamental mode the differences are always smaller than $4,8\%$, they can become as great as $12\%$ for the sixth overtone. As for the waveform we show that, although the first oscillations follow almost the same profile for all the values of $\xi$ considered, increasing values of $\xi$ imply that the perturbations are less suppressed, which at the physical level translates into longer ringdown signals and results in greater differences compared to the classical Schwarzschild case.

The comparison with other regular black hole metrics highlights the same key points, that are the greater sensitivity to the spacetime structure of higher overtones and of the length of the ringdown. The scalar perturbations of the Bonanno-Reuter regular black hole, which is the counterpart of the Schwarzschild solution in the context of asymptotically safe gravity, do not differ much from the ones of a non-singular collapse with a specific value of $\xi$. As for the Hayward metric, which is a minimal model for a generic regular black hole metric, the differences are more significant; in particular, the ringdown waveform decays faster than a Schwarzschild black hole with the same mass, rather than slower as in the non-singular collapse model.

Overall, these results provide a framework to test stability of solutions and horizon geometries dependence. It could be important also for future tests extending the analysis to all the component of gravitational perturbation they would help, thanks to the future detection of the new generation of detector, to distinguish between one model and another.

\bibliographystyle{ws-procs961x669}
\bibliography{ws-pro-sample}

\end{document}